# Parametric frequency fusion by inverse four-wave mixing

**Thibaut Sylvestre**
*Laboratoire d'Optique P.M Duffieux, U.M.R CNRS/Université de Franche-comté n°6603*
*25030 BESANCON cedex, FRANCE*
*fax: 33(0)3.81.66.64.23*
email: thibaut.sylvestre@univ-fcomte.fr

*Abstract*
This work reports the experimental observation of a new type of four-wave mixing in which frequency-degenerate weak signal and idler waves are generated by mixing two pump waves of different frequencies in a normally dispersive birefringent optical fiber. This parametric frequency fusion is what we believed the first experimental evidence of inverse four-wave mixing.

**Introduction**

A well-known kind of parametric wave mixing in nonlinear optics reads as $2\omega_0 \rightarrow (\omega_0+\Omega)+(\omega_0-\Omega)$, that is, the four-wave photon interaction of a central frequency wave at frequency $\omega_0$ with a pair of up-shifted (anti-Stokes) and down-shifted (Stokes) waves at frequencies $(\omega_0+\Omega)$ and $(\omega_0-\Omega)$, respectively[1,23]. This phenomenon is fully related to another well-known nonlinear process called modulational instability (MI), whereby weak modulations of a continuous-wave are spontaneously amplified, that leads to the generation of soliton-like periodic pulse trains in the time domain, or equivalently, to the generation of soliton-like beam arrays in the spatial domain in a 1+1D configuration[5,6]. Although FWM and MI have been independently studied in the past for both fundamental interests and their numerous potential applications; the link between both is now well recognized by several authors[1].

Four-wave mixing is achieved when the four interactive waves fulfill a phase-matching condition that ensures the energy-exchange process between the waves all along the propagation in the nonlinear medium. For instance, in optical fibers, four-wave mixing has been observed under various phase-matching conditions in both normal and anomalous dispersion regime. In most cases, however, four-wave mixing process is degenerated, i.e., two identical pump photons are annihilated and a pair of up-shifted (anti-Stokes) and down-shifted (Stokes) photon are generated. This unilateral energy transfer opens the question as whether it is possible to achieve *the inverse process of usual four-wave mixing in nonlinear media*. As recently suggested by C.F. McCormick *et al.*[7], the inverse process should be also degenerated and should satisfy the energy transfer relation $(\omega_0+\Omega)+(\omega_0-\Omega) \rightarrow 2\omega_0$.

In this work, we report what we believed to be the first experimental evidence of the inverse of modulational instability or four-wave mixing, by mixing two pump waves of different frequencies in a normally dispersive birefringent optical fibers and using a cross-polarization phase matching technique[8]. Experimental results show that a weak-wave formed by a frequency-degenerated signal and idler pair spontaneously grows through four-wave mixing, which results in the parametric frequency fusion of the two pump waves. As a matter of fact, this *parametric wavelength fusion* is an intrinsic property of the reversible behaviour of modulational instability recently demonstrated in optical fibers by G. Van Simaeys *et al.*[9] and also known as the Fermi-Pasta-Ulam recurrence.

Let me introduce the phase-matching technique involved in the present study that leads to vector modulational instability[10] (schematically depicted in Fig. 1). In the normal dispersion regime of an optical fiber ($\beta_2 > 0$), phase-matching for inverse four-wave mixing is achieved when the phase mismatch induced by the linear birefringence compensate for both the pumps-induced nonlinear phase shift (cross-phase modulation) and the phase mismatch due to chromatic dispersion. In the case of a highly birefringent fiber, inverse four-wave mixing will occur for two pump waves of different frequencies orthogonally polarized along the fast and slow axis of the fiber, respectively. The nonlinear interaction between the two frequency-detuned pump photons will produce a pair of signal-idler photons of equal frequency and orthogonal polarizations. The expected weak signal will thus present two polarizations components.

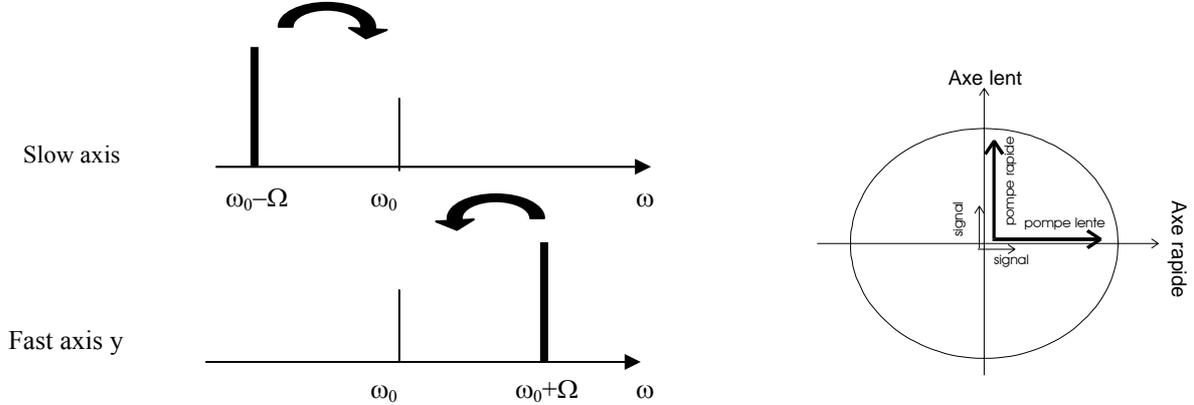

*Figure 1: Scheme of the inverse four-wave mixing process and cross-pump divided phase-matching in a normally dispersive highly birefringent fiber.*

In such a configuration, the phase-matching relationship between the propagation constants is given by

$$\Delta\beta = \beta_{\omega_0 x} + \beta_{\omega_0 y} - \beta_{(\omega_0-\Omega)x} - \beta_{(\omega_0+\Omega)y}$$
$$= -\frac{\Delta n \Omega}{c} + \beta_2 \Omega^2 - \gamma(P_{(\omega_0-\Omega)x} + P_{(\omega_0+\Omega)y}) = 0, \quad (1)$$

where $\Delta n = |n_x - n_y|$ is the fiber birefringence, $2\Omega$ is the frequency detuning between the pumps, $\beta_2$ is group-velocity dispersion coefficient, $\gamma$ the nonlinear coefficient, and $P_{(\omega_0-\Omega)x}; P_{(\omega_0+\Omega)y}$ are the pump powers. Note the opposite sign for the last term of Eq.(1) in comparison with respect to the usual four-wave mixing process. The optimum frequency detuning $2\Omega$ between the two pumps for a perfect phase-matching is given by

$$2\Omega = \frac{1}{\beta_2}\left[\frac{\Delta n}{c} + \sqrt{\left(\frac{\Delta n}{c}\right)^2 + 4\beta_2\gamma(P_{(\omega_0-\Omega)x} + P_{(\omega_0+\Omega)y})}\right] \quad (2)$$

By inserting in Eq.(1) the values of parameters used in our experiment in the visible domain : $\lambda=532$ nm, $\beta_2=6.10^{-26} s^2 m^{-1}$, $\Delta n=5.5.10^{-4}$, $\gamma=53 W^{-1} km^{-1}$, the optimum frequency detuning $2\Omega$ for achieving inverse four-wave mixing must be equal to 8.5 THz (8 nm@532 nm), for a total peak pump power $P_{(\omega_0-\Omega)x} + P_{(\omega_0+\Omega)y} = 100$ W.

If we now consider the scalar case in the anomalous dispersion regime ($\beta_2<0$), the phase-matching can be satisfied without birefringence, when the pump-induced nonlinear phase shift (cross-phase modulation) compensate for the phase-mismatch induced by chromatic dispersion, all the interactive waves being linearly co-polarized. Thus, a well-defined condition allows for the fusion of two pump photons of different frequencies, and to the generation of a pair of signal/ilder photons both with equal frequency and state of polarization. The phase-matching condition is given by

**Experiment**

The experimental setup is shown in Fig. 2. The first pump wave ($P_x$) is a 35-ps light pulse delivered by a frequency-doubled mode-locked Nd:YAG laser operating at 10 Hz. Its spectral width is 0.028 nm at 532 nm wavelength. The second pump pulse ($P_y$) is provided by an optical parametric generator (OPG) pumped by the third harmonic of the Nd:YAG laser at 355 nm. The second pump pulse is temporally stretched by a diffraction grating from 18 ps to 40 ps and its wavelength is tunable within the visible range 420-680 nm. We have carefully adjusted its wavelength at 524 nm in agreement with the above theoretical calculations ($2\Omega/2\pi=8.5$ THz).

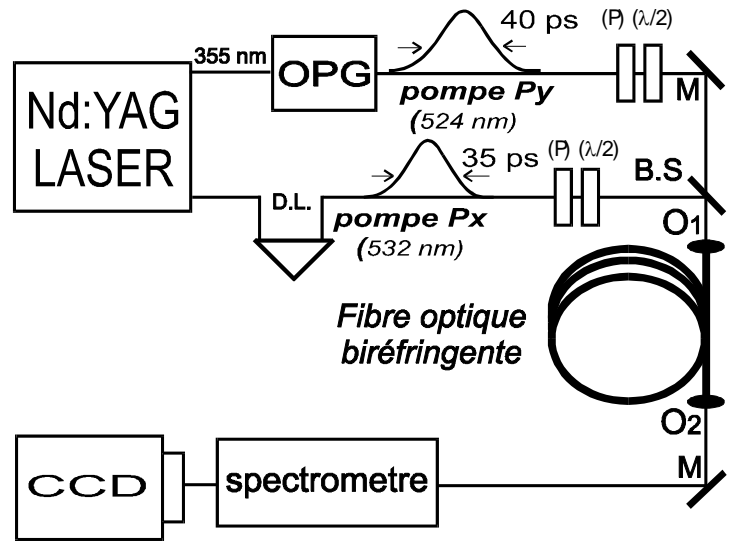

*Figure 2:* Experimental setup for the observation of inverse four-wave mixing. M: Mirrors, B.S: beam splitter (50/50), (O1,O2): microscope objectives, P: Glan polarizer, $\lambda/2$: half waveplate, D.L.: Delay line.

The two pump pulses are synchronized temporally at the entrance of the optical fiber with an optical delay line and a streak camera (time resolution of 5 ps). The single-mode polarization maintaining fiber (Newport Corp. FSPA-10 model) has a core diameter of 1.9 µm, a cladding thickness of 125 µm and a birefringence of $5.5.10^{-4}$. The fiber length was 4 m so that the pulse walk-off between the pumps remains negligible in the energy-exchange process. The pumps $P_x$ and $P_y$ are polarized with half waveplates on the slow and fast axis of the fiber, respectively. The input pump power is also measured pulse per pulse and is calibrated using a joulemeter. Finally, the output light was spectrally analysed by a Littrow grating spectrometer (0.5 m, 600 lines/mm) with a spectral resolution of $4.10^{-2}$ nm. The experimental spectra are recorded pulse per pulse with a single-shot CCD camera.

Figs. 3.(a)-(b) show respectively the image and the profil of the output spectrum for a total input peak power of 200 W. As can be seen, a small-signal at 528 nm is generated, exactly at the mean wavelength between the two pumps. The two pumps parametrically interacted to generate a weak-wave by inverse four-wave mixing.

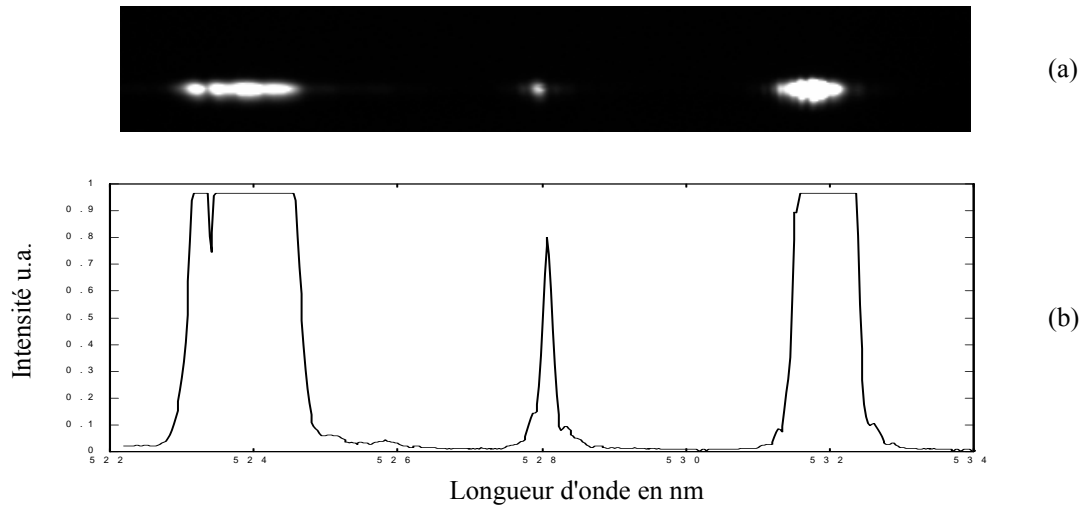

*Figure 3:* Output spectrum (a) and corresponding profil (b) showing the parametric wavelength fusion of two pumps waves. $P_x+P_y=200W$.

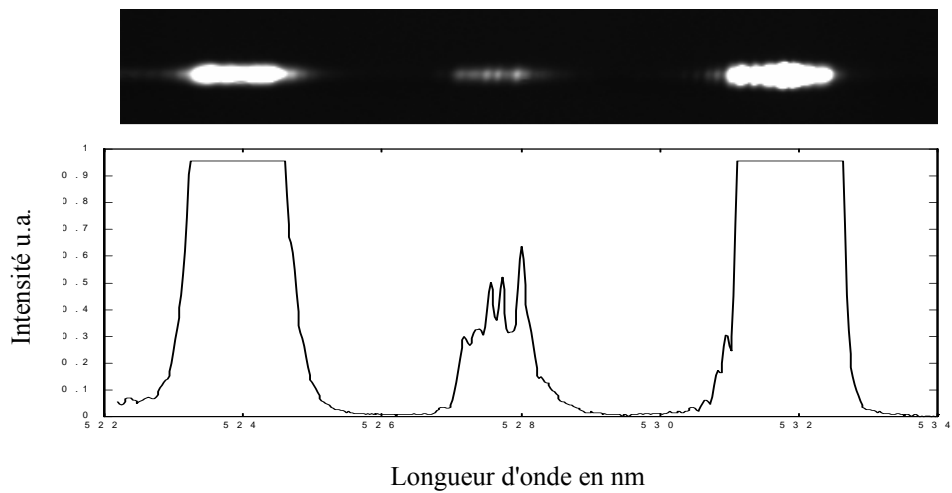

*Figure 4:* Output spectrum (a) and corresponding profil (b) showing the parametric wavelength fusion of two pumps waves. $P_x+P_y=300W$.

Note that the spectral broadenings of the pump pulses are due to self and cross-phase modulation and are relatively significant in the picoseconde regime. Indeed, the strong nonlinear phase shift of γPL=13 π radians induces more than 13-fold increase of the input pulse spectrum. Figs. 4.(a)-(b) illustrate the output spectrum with a peak power of 300 W. The small signal between the two pumps is more intense than in Figs. 3(a)-(b) and presents a spectral broadening also due to self and cross-phase modulation with the pump pulses.

In conclusion, I have designed a simple experimental arrangement using an optical fiber that allowed me to observe what I believed to be the inverse degenerated four-wave mixing process, i.e., two pump waves of different frequencies will parametrically fusion to generate a single-wave from noise at the exact mean-frequency. This has been achieved by using a cross-polarization phase-matching technique in the normal dispersion regime of a highly birefingent fiber. The experimental measurement of the ideal pump wavelength detuning is in excellent agreement with the analytical result. I suggest that this effect should be also experimentally observable in the anomalous dispersion regime of optical fibers. Inverse four-wave mixing results in the parametric fusion of frequencies and can be related to the reversible behavior of modulational instability, when energy goes back from the parametric Stokes and anti-Stokes waves to the initial pump wave. In addition, the generation of a weak-wave between the pumps results in a drop of contrast of the periodic modulation induced by the beating note of the two pumps. This effect may find significant applications in noiseless optical parametric amplification operating in the phase sensitive configuration as the signal and idler waves are frequency-degenerate. The phase sensitivity can be thus directly induced by controlling the pumps phase.